# Automated detection of wild and cultivated plants in citizen science records

Christophe Botella[1*], Joy Mangachena[2], Sjirk Geerts[2], Hervé Goëau[3], Luke J. Potgieter[4], David M. Richardson[4], Cang Hui[5,6]

1: Inria, LIRMM, University of Montpellier, Montpellier, France
2: Centre for Invasion Biology, Department of Conservation and Marine Sciences, Cape Peninsula University of Technology, Cape Town, South Africa
3: AMAP, Univ Montpellier, CIRAD, CNRS, INRAE, IRD, Montpellier, 34980, France
4: Centre for Invasion Biology, Department of Botany & Zoology, Stellenbosch University, Stellenbosch 7602, South Africa
5: Department of Mathematical Sciences, Stellenbosch University, Stellenbosch 7602, South Africa
6: NITheCS, African Institute for Mathematical Sciences, Cape Town 7945, South Africa

*Mail: christophe.botella@inria.fr

**Highlights:**

- We developed a decision framework to classify wild and cultivated plant records.
- Cultivated plants contaminate 21.4% of Research Grade iNaturalist records.
- Automated detection reached 87% accuracy, better than iNaturalist cultivation tag.
- Predictions on all records identified known naturalization hotspots in Cape Town.

**Abstract:** Citizen science plant occurrences may help the early detection of naturalisations of alien species, yet they are affected by misclassifications of cultivated plants as wild occurrences. To differentiate cultivated from wild observations, we proposed a decision framework for the manual annotation of records. We annotated this cultivation status for 1,491 iNaturalist records in Cape Town, comprising 12 invasive alien plant species. To automate the classification, we trained algorithms on annotated records based on 39 predictors from split into 4 sets: (i) basic record and observer features, (ii) visual features extracted from images, (iii) geographic features, and (iv) iNaturalist observer statistics. 11 Classification algorithms were trained on 8 combinations of predictor sets, to assess predictors importance, yielding 88 models evaluated with cross-validation. 25.2% of all annotated records were cultivated, and 21.2% among Research Grade records, while 24.8% of annotated records were doubtful. Automated classification yielded an accuracy in distinguishing cultivated from wild records reaching 87%, improving over the iNaturalist cultivation tag (74%), and the accuracy was robust when predicting on new species (85%). Model performance was overall highest when combining iNaturalist-specific, geographic and visual predictors, but visual predictors alone performed better for certain species. Our results highlight the potential to automatically distinguish cultivated and wild-growing plant records to support invasion risk assessments and sanitise ecological analyses currently contaminated by cultivated records. Our framework is transferable to other crowdsourcing platforms, even though the manual annotation capacity can limit transferability.

# 1. Introduction

Biological invasions are one of the main drivers of ongoing biodiversity loss globally (Roy et al., 2024). In the process of invasions, naturalisation is a key step characterised by the ability of an introduced species to form self-sustaining populations (Richardson and Pyšek, 2012). Early detection of plant naturalisation is central to preventing large-scale invasions that often result in high management costs (Ibanez et al., 2009; Venette et al., 2021; Mologni et al., 2024). However, naturalisation events are difficult to monitor as they require intensive and repeated sampling (Kalwij et al., 2014; Dodd et al., 2015) and, by definition, the ability to discriminate between wild and cultivated plants, which often occur in close proximity (Seitz et al., 2022). Global crowdsourced biodiversity databases provide a unique opportunity to address this challenge, providing free, dense and regularly updated observations in populated areas (Encarnação and Morais, 2021), which host a myriad of horticultural plant species and are probably the most important source of plant invasions (Van Kleunen et al., 2018). Despite the huge potential of citizen-science presence-only records (hereafter CS-PO), they are highly contaminated by records of cultivated plants, limiting their usefulness for early detection and, more broadly, for biogeographical studies (Encarnação and Morais, 2021; Pocock et al., 2023). No international crowdsourcing platform provides a robust framework to differentiate between wild and cultivated plants, partly because the concept of cultivation can be blurry and context dependent. For instance, cultivation can be understood strictly in an agricultural sense, or more broadly to include gardening and ornamental activities; it can encapsulate or not a diversity of practices promoting a plant; and it can encapsulate or not plants that are unintentionally favored by these practices (e.g. weeds). Hence, we primarily lack a definition that is clear and useful for studying naturalisations, operational criteria for determining the cultivation status of plant records, and also transferable and scalable methods to assess it automatically and unlock the potential of crowdsourced databases for ecological research and invasion monitoring.

Urban areas - here defined as areas dominated by built development and high-density human infrastructure - experience a greater influx of alien species introductions (both accidental and intentional) than rural or natural areas (Potgieter et al. 2024). Human activities such as increased disturbance, resource supplementation, and transport networks provide many opportunities for alien species to establish, proliferate, and spread, thereby facilitating invasions within and around urban areas (Kowarik 2011; Hui et al., 2017; Cadotte et al. 2017; Potgieter and Cadotte 2020). However, we know surprisingly little about alien plant species distribution and dynamics in urban areas compared to natural areas (Gulezian and Nyberg, 2010; Lososova et al., 2012; Gaertner et al., 2017), but see Alston and Richardson, 2006; Afonso et al., 2020. In particular, it is difficult to reconstruct invasive plant population dynamics in urban areas due to the difficulty of distinguishing between wild and cultivated populations (Botella et al., 2022; Li et al. 2025). More broadly, much uncertainty remains on how to characterise urban infrastructure and land use as habitats shaping biodiversity responses (Li et al., 2019), which has arguably contributed to the fact that urban areas are rarely accounted for in conservation planning (Cooper et al., 2007).

With the ever-increasing volume of data from citizen-science projects, more information exists on alien plants in urban areas than ever before (Spear et al., 2017). Although the sampling is typically biased towards densely populated areas (Ward, 2014), this bias can be advantageous when monitoring urban biodiversity, its response to the physical anthropogenic environment, its spatial configuration, and its implications thereof for conservation (Cooper et al., 2007; Altwegg et al., 2017; Li et al., 2019). Citizen-science data also partly overcomes the problem of limited access to private land (Spear et al, 2017) and can directly inform the management of biological invasions in urban areas (e.g., Crall et al., 2010; Potgieter and Richardson, 2024; Potgieter et al., 2024a, b). Most studies comparing CS-PO data with professionally collected data have found broad agreement or similar trends (Aceves-Bueno et al. 2017), suggesting that CS-PO data can enhance our understanding of alien and invasive plants in urban environments (Isaac et al., 2014). However, their usefulness for alien plant monitoring in urban settings remains largely unexplored.

The first barrier to a more generalised use of CS-PO data pertains to their inherent sampling biases (e.g., Burgess et al., 2017), which often reduce confidence in their application and limits their uptake in ecological research (Pocock et al., 2017; 2023; Encarnação and Morais, 2021). Volunteer contributors of CS-PO tend to focus on certain taxonomic groups (DiCecco et al., 2021; López-Guillén et al. 2024), have questionable identification skills (Crall et al., 2011), and sample more unevenly across space and time than professional sampling schemes (Boakes et al., 2010). Sampling bias problems have been extensively discussed in the literature, and many statistical techniques have been developed to address them (Bird et al., 2014). However, one other important limitation that to our knowledge has not been explicitly addressed, is the need to discriminate between cultivated and wild (free-growing, including established or naturalised alien) plant records, which often requires consideration of multiple sources of information such as geolocation, photographic evidence and characteristics of the observer. This is one of the main challenges for generalising the use of crowdsourced biodiversity databases in monitoring the distribution and naturalisation of alien plant species. Indeed, crowdsourcing platforms (e.g. iNaturalist, Pl@ntNet) contain large numbers of cultivated plant records, which can bias species distribution models by overestimating the suitability in areas where plants are nurtured by humans (see e.g. Botella et al., 2018; Atsumi et al., 2024). Certain platforms, notably iNaturalist, provide observers with the option to tag their records as "*cultivated*". For an alien species, the absence of this tag might be assumed to imply the presence of a mature individual that has survived without cultivation. This would indicate that the species has reached stage C1 in the introduction-naturalisation-invasion continuum (Blackburn et al., 2011) and might be already self-perpetuating in the wild (stage C2 or naturalised s.s., Richardson et al. 2000). However, this inference is often unreliable as many records of cultivated species are not tagged as "cultivated".

Determining the cultivation status of a plant in the field is often difficult, particularly for invasive species, which frequently grow spontaneously close to, around, or among cultivated populations (Kowarik, 2005; Palit et al., 2024). In this context, it is important to distinguish "planted" and "cultivated" plants: signs of cultivation (e.g., weeding, irrigation, pruning or fertilisation) are generally easier to detect than signs of intentional planting. We focus here on the former signs of cultivation status as they are particularly relevant to the study of naturalisation, given that cultivation practices often play a key role in alleviating the stresses and negative interactions that may otherwise limit a species' ability to grow (Potgieter and Cadotte, 2020). However, evidence of cultivation status may be lacking for various reasons, and it is therefore important to account transparently for this uncertainty when distinguishing cultivated from wild plants.

Our aim is to propose a transparent and evidence-based decision framework to distinguish cultivated from wild and doubtful (uncertain cultivation status) plants in crowdsourced records, and to assess the potential of machine-learning methods to automate this task in a transferable way (Yates et al., 2018). We propose criteria to determine the cultivation status of a plant from the available information (photos, coordinates, observer tags, profile and comments; also, geographic covariates). Cape Town in South Africa, and its peri-urban areas, form an ideal study area with high iNaturalist uptake (Richardson & Potgieter 2024), hosting a diverse alien flora (Richardson et al., 1996, Gaertner et al., 2016) sharing species with many cities around the world, while its urban properties are known to facilitate plant naturalisation (Potgieter et al. 2020, Afonso et al., 2020). We use our decision framework to annotate 1,491 iNaturalist records in Cape Town as a case study dataset to train and test machine-learning methods for classifying plant records based on 39 selected predictor variables split into 4 predictor sets: basic record and observer features, visual features extracted from images, geographic context features, and iNaturalist-specific observer statistics. This allowed us to compare the reliability of the machine learning methods to the observer-provided cultivation tag in iNaturalist for classifying the cultivation status. We finally illustrated the predictions of the best algorithm on all 99,408 trusted iNaturalist plant records in the Cape Town area to identify potential naturalisation hotspots.

## 2. Methods

### a. Study area and species list

Cape Town was chosen as a study area primarily because its data have good potential to support transferable of our method elsewhere: the city is very well sampled in iNaturalist, covering a large diversity of alien plant species with a taxonomic composition similar to that of many other cities around the world (Potgieter et al., 2020). Widespread plant invasions threaten the city's rich native biodiversity. Indeed, Cape Town is particularly vulnerable as it is located within a biodiversity hotspot – the Cape Floristic Region - the smallest but most biologically diverse plant kingdom – with high native species richness and levels of endemism (Holmes et al., 2012; Myers et al., 2000). Cape Town hosts 52% of the country's critically endangered vegetation types (Rebelo et al., 2011). With a population of 4.8 million people, which is growing more rapidly than any other southern African city (Boraine et al., 2006), the municipal area has been rapidly transformed for urban development (26% of the municipality area) and agricultural use (35%), while the rest of the area remains under natural or semi-natural vegetation in mountainous areas mainly within Table Mountain National Park (Richardson et al., 1998; Holmes et al., 2012). With hundreds of alien plant species introduced in the past few centuries (Richardson et al., 1996; Foxcroft et al., 2008; Geerts et al., 2017), urbanisation has likely facilitated the invasion processes observed in the area (Potgieter et al., 2020). We delimited our study area more precisely to the five quarter-degree cells shown in Figure 1, which correspond to the Atlas cells of the BODATSA botanical record database (extracted from GBIF.org, 14 November 2022) that best overlap Cape Town's municipal area. This enabled us to automatically derive a reliable native plant species list, which was complemented with a recent inventory of alien plant species (Rebelo et al., 2025) that was taxonomically matched to the GBIF backbone taxonomy using the R package rgbif.

### b. iNaturalist dataset

We extracted all iNaturalist records located in the study area using the R package rinat (extraction from 12/08/2022). iNaturalist taxa were matched with the GBIF backbone taxonomy using the R package rgbif to filter the names that did not appear in the full species list (native and alien) for the study area. We also excluded records located in the ocean and inland water bodies based on the GlobCover land cover raster (ESA GlobCover 2009 Project), and those for which the positional accuracy was not precise enough (uncertainty >100m). Finally, we retained only records for which identification was trustable by applying the same rule as for Research Grade records, i.e. the plant is identified by at least two users including the observer with at least two-thirds agreement on the final species name, except that we also kept trusted records tagged as “cultivated”. Thus, records tagged as “cultivated” by the observer were kept only if their identification was trustable. Our complete iNaturalist dataset comprised 99,408 records with trustable location and identification (Figure 1), of which 1,654 were tagged as “cultivated”. These records represented 2,118 species, of which 381 were alien.

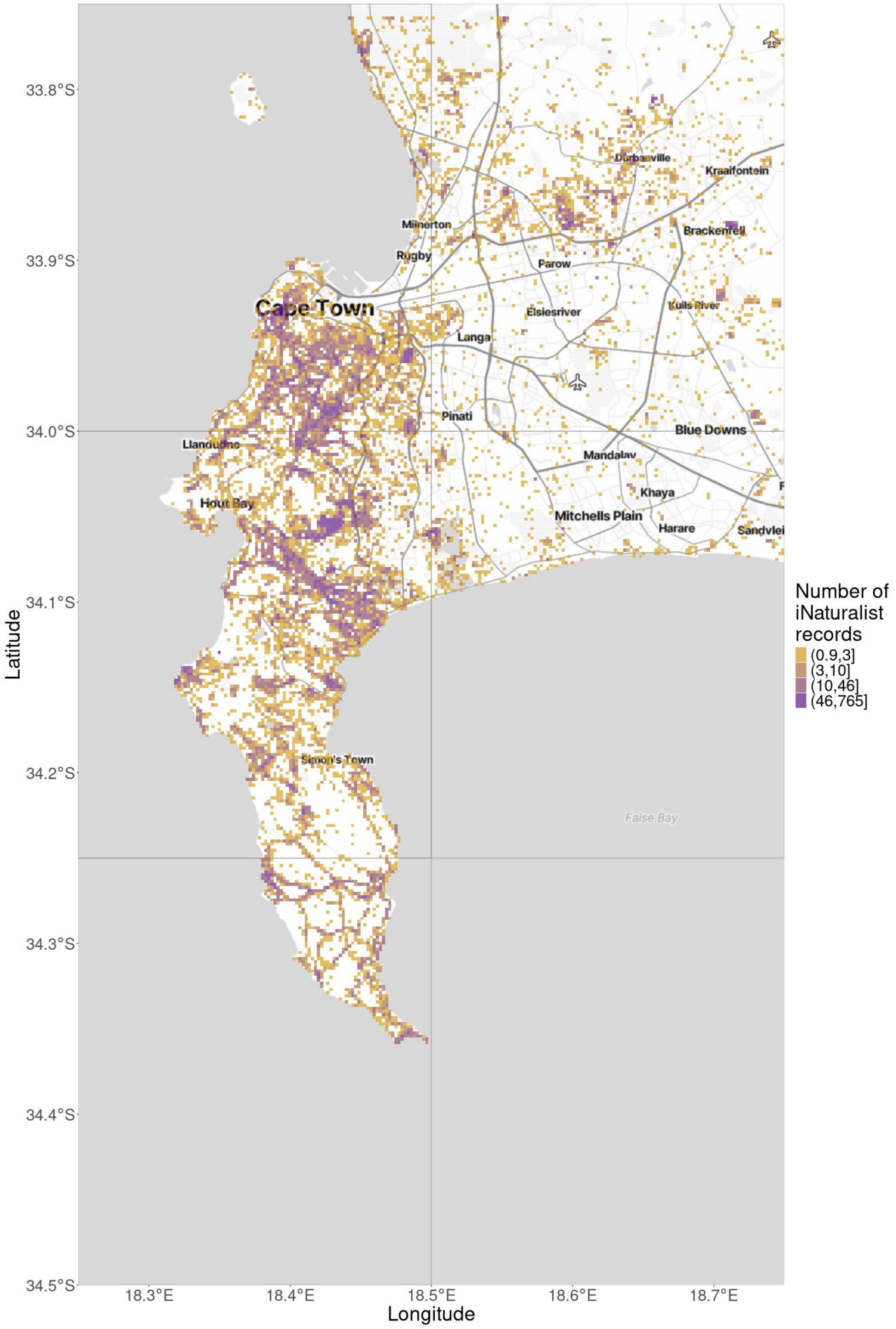

**Figure 1.** Cape Town study area and the 99,408 trusted iNaturalist plant records until August 2022. These trusted records fulfil the Research Grade criteria but include those tagged as “captive or cultivated” by the observer; they have a geographic uncertainty <100m and are not located in water bodies. Our study area comprised the five quarter-degree cells delimited by the solid black lines. Colours indicate the number of records per 200 m pixel (i.e. 4 ha area). Background maps are based on Stadia Maps, Stamen Design, OpenMapTiles and contributions to OpenStreetMap.

### c. Cultivation status of plants from iNaturalist records and dataset annotation

Explicit criteria to assess the cultivation status of a plant specimen are surprisingly rare in the literature. A range of situations are ambiguous, making it difficult to define a precise limit between wild and cultivated plants. Seitz et al. (2022) sampled gardens across Berlin for wild and cultivated plant species and considered plants as cultivated on the basis of “traces of cultivation (cultivation in rows or beds, promotion by weeding, component of typical ‘wildflower’ seed mixtures)”. Such criteria need to be refined to be applied to the diverse situations encountered among iNaturalist records, but they highlight the promotion of plants by weeding as an important sign of cultivation. While proving that a plant was sown or planted by humans is often impossible without contextual knowledge, some evidence may prove that its growth was promoted by humans relative to its surrounding competitors through, e.g., weeding, irrigation, pruning of secondary branches.

In general, we define as *cultivated* any plant that has been actively selected or promoted by humans to the detriment of other plants that would have otherwise competed with it for natural resources (light, soil or water not brought by humans). This may notably include direct descendants of surrounding cultivated plants. Conversely, any plant that survives without human intervention, or in spite of attempts to restrain its growth, and that visibly competes with other plants for natural resources is considered *wild*. Yet, it may be impossible to discern from the accessible information whether an observed plant was cultivated or wild, in which case its status is *doubtful*. Unfortunately, the above definitions are often not suited to determine the actual cultivation status from the evidence available in a digital plant record and thus require further refinement and assumptions.

For this study, we designed transparent criteria to annotate any iNaturalist plant record as either cultivated, wild or doubtful. Even though these criteria are only a proxy for the real cultivation status, they are practical for the manual annotation of iNaturalist records and classification errors should remain marginal for the envisioned contexts. We considered criteria based on evidence found in the geolocation, images, or observer comments. First, we look for strong evidence of cultivation. If none is found, we look for strong evidence of wild status. Finally, if none is found, the status is considered doubtful. We list below the criteria for evidence of cultivation. Any of these leads to the record being classified as *cultivated* in our annotation. The images of a record show signs of the following directly around the plant:

- **Soil culture** (soil perturbed by a human tool), **mowing** or **weeding** or other evidence of **active cultivation** to reduce competition imposed by other plants (e.g. mowing, weeding)
- **Irrigation** (e.g. pipes)
- **Human pruning** of the plant branches based on scars along the trunk or branches, but excludes large cuts at the base of the trunk or any other pruning showing past attempts to eradicate the plant; e.g. example

- The plant being **inside** a building, a greenhouse, planted in a pot, or any other human-made infrastructure; e.g. example 1, example 2. This criterion excludes plants growing in outdoor gardens or roofs, which may be wild.

In the absence of strong evidence of cultivation, if any of the following criteria is fulfilled, the record is classified as *wild*:

- **Other plant species grow freely close to the plant** (≤1 m) and the plant is not inside a cropland, pasture or garden. This land use constraint does not apply if the plant is a known weed of the current land use, otherwise we would filter many weed records.
- The plant is **in a bush, an edge or forest, a wetland, on a river bank**.
- The plant is **in a protected area** with no human settlements in the surroundings.
- The local vegetation is not managed by humans (based on e.g. land use data, or visibly no human infrastructure or activity, except roads and trails).

In the absence of such evidence, the plant is classified as **doubtful**, unless the notes indicate that the plant is cultivated or wild, in which case the record is classified accordingly. This hierarchy of criteria is summarised in Figure 2**.**

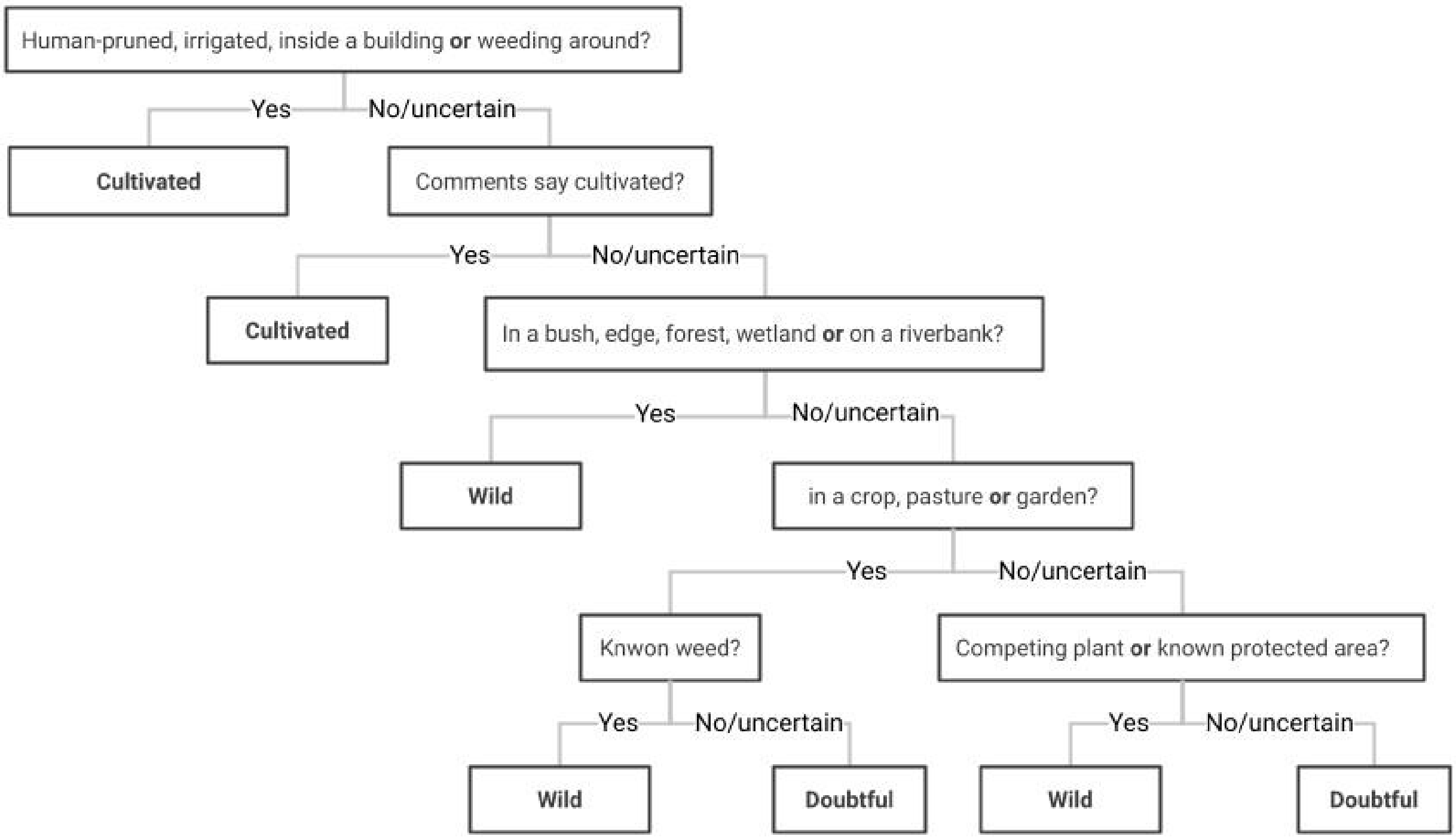

**Figure 2.** A decision tree used to classify a plant record as cultivated, wild or doubtful in our manual annotation.

We used the above criteria to manually annotate all trusted iNaturalist records of 12 invasive alien plant species in our study area, yielding a total of 1,491 annotated records, among which 744 (49.9%) were wild, 376 (25.2%) were cultivated and 371 (24.9%) were doubtful. These species (*Melaleuca viminalis*, *Centranthus ruber*, *Duranta erecta*, *Echium candicans*, *Ficus carica*, *Hypericum canariense*, *Phoenix canariensis*, *Psidium guajava*, *Schinus terebinthifolia*, *Solanum* *lycopersicum*, *Spartium junceum*, *Tecoma stans*) were chosen because they had a significant number of records and are both cultivated and present in the wild (e.g., Geerts et al., 2013, 2017; Afonso et al., 2022).

Example images of annotated records are provided for each of the three statuses (**Figure 3**), and an interactive map of all annotated records is provided in a Shiny web application: https://chrisbotella.shinyapps.io/managed_vs_wild_annotated/

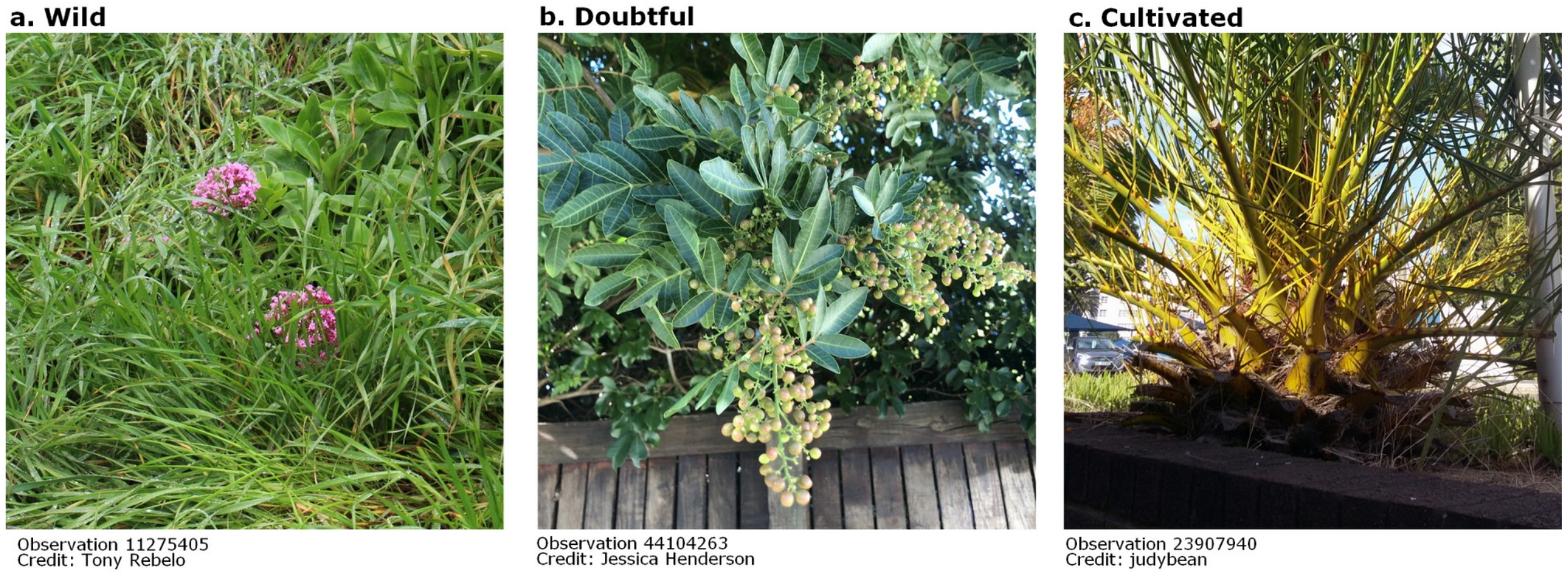

**Figure 3.** Three iNaturalist records that we manually annotated. (a) A wild *Centranthus ruber* plant, (b) a doubtful *Schinus terebinthifolia* and (c) a cultivated *Phoenix canariensis*. (a) is wild because it is in direct competition with herbaceous plants, (b) is doubtful because there is no evidence for its wild or cultivated status, and (c) is cultivated because its leaves have been pruned and the grass around is mowed.

### d. Input variables for automatic cultivation status prediction

We extracted various features associated with each iNaturalist record, including the picture, observer profile and geographic context, in order to test whether the cultivation status may be predicted automatically. We selected 39 candidate predictor variables categorised into four sets (basic, iNaturalist-specific, geographic and visual, see **Table 1**).

Basic predictor variables are observer- or record-level statistics not specific to iNaturalist (positional_accuracy, lnObs, lnSp in **Table 1**) and could be derived from other data sources, hence transferable. iNaturalist-specific predictor variables include the presence of the "*cultivated*" tag submitted by the iNaturalist observer (iNat_tag), two identification statistics (nReviews, agreeRate) and two indicators of the observer use of the "*cultivated*" tag (tagged, tagFreq). These variables are not necessarily transferable as they depend on specific iNaturalist functionalities. The four observer level input variables (lnObs, lnSp, tagged, tagFreq) were computed from all iNaturalist records in the study area (even the "non-trusted" ones, i.e. those where the geolocation or identification was uncertain).

The geographic variables related to distance to watercourses, protected areas and land parcels were extracted from the City of Cape Town Open Data Portal (https://odp-cctegis.opendata.arcgis.com/, accessed on 15th July 2022). Distances to natural and artificial watercourses (dToRiverM, dToCanalM) were computed from the watercourse segments of the Open Watercourse dataset considering the categories "*Canal Lined*", "*Channel Open*", "*Canal

*Composite*", "*Gap Connector/Flow Path*" and "*Stream Extension*" for the artificial ones, and the category "*Natural Rivers/Streams*" for the natural ones. We used all polygons from the Biodiversity Network dataset of Cape Town to determine whether each record was in a protected area (protecArea). The simplified land-parcel status of each location was obtained from the City of Cape Town Cadastre. This polygon database has an expected polygon delimitation uncertainty below one metre, and can differentiate, for instance, a street from an adjacent private garden or park as illustrated in Figure S1 of Appendix 1. The polygon categories considered residential (parcel_resid) were indices 1-6 in the category "General residential" and index 1 in category "Single residential". Business areas (parcel_business) aggregated three categories: "*General Business*", "*Local Business*" and "*General Industrial*"). Roads (parcel_road) had only one category "*Transport 2: Public Road and Public Parking*"), as did agricultural areas (parcel_agri, "*Agricultural*") and open areas (parcel_open, "*Open Space*").

For the land cover, taken from Globcover ([ESA GlobCover 2009 Project]), we simplified the 73 original categories into seven categories (Natural Wooded Land, Planted Forest, Shrubland, Residential, Urban Vegetation, Built-up and Other, see Table S1 of Appendix 2) to ensure sufficient representation of each simplified category in our dataset. This input variable is transferable spatially as Globcover provides the land cover at 300m resolution globally.

We also included visual input variables describing the content (foreground and context) of the first image of each record, obtained from a dedicated image recognition algorithm. For the image foreground, i.e. objects in the image focal area, we had 18 visual classes including various plant organs, animal taxonomic groups, mushrooms or landscape (if the picture has a very broad focus). For the context, i.e. the rest of the image including its background, the values of the four input variables correspond to the probabilities that the context belongs to four categories: anthropized (e.g., built up surfaces, walls, lighting, fences), presence of human body (e.g. hand, head or other body parts in the background), wild (no traces of humans or human-made infrastructures), uniform background (e.g. paper). The predicted class probabilities for the image foreground and context both come from the same bi-modal classification deep learning model, initially developed to filter irrelevant images submitted to Pl@ntNet. This model is a BEiT vision transformer (Bao et al., 2021) trained on thousands of manually labelled images of the Pl@ntNet database, and available with code for reproducibility (Goëau et al., 2026).

**Table 1.** The 39 input variables for automatic cultivation status classification and their type: Record/observer level (iNaturalist specific or not), geographic and visual.

| **Input variable name** | **Input set** | **Description** |
|---|---|---|
| Positional_accuracy | Basic | Uncertainty of the record geolocation in metres |
| lnObs | Basic | Logarithm of the total number of records collected by the observer (proxy of observer experience) |
| lnSp | Basic | Logarithm of the total number of species collected by the observer (proxy of observer experience) |
| iNat_tag | iNaturalist specific | Whether the plant record was tagged as “cultivated” by the observer in iNaturalist |
| nReviews | iNaturalist specific | The number of taxonomic identifications of the record in iNaturalist (observer + identifiers). |
| agreeRate | iNaturalist specific | The proportion of the record identifications agreeing with the final identification |
| tagged | iNaturalist specific | Whether the observer has already tagged at least one record as “cultivated” in iNaturalist |
| tagFreq | iNaturalist specific | The proportion of the observer records tagged as “cultivated” in iNaturalist |
| protecArea | Geographic | Whether the location is in a protected area based on Biodiversity Network dataset of Cape Town municipality |
| dToRiverM, dToCanalM | Geographic | Distance to natural and artificial watercourses, respectively, in metres based on the Cape Town Open Watercourses dataset |
| Parcel_ ... resid, business, road, agri, open | Geographic | Whether the record location belongs to various land parcel simplified categories according to the city of Cape Town Cadastre including: residential area, business area, road, agricultural field or open area. One location may belong to several categories. |
| simple_lc | Geographic | Simplified land cover category (7 classes, see **Table S1 of Appendix 2**) based on GlobCover raster (resolution ~300m) |
| Visu_for_ ... humanmade/ landscape/ bark/ branch/ bud/ flower/ fruit/ habit/ leaf/ roots/ seeds/ amphibian/ reptile/ bird/ mammal/ | Visual | Predicted class probability of the record image **foreground** (focus of the photograph) among 18 possible classes including human-made object, landscape, various plant organs, various animal groups or mushrooms. |
| fish/ invertebrate/ mushroom | | |
| Visu_ctx_ … anthropized/ intothewild/ uniform_background_macro_dissection/ with_human_body_presence | Visual | Predicted class probability of the record image **background/context** (surrounding the focus of the photograph) among 4 possible classes: A human transformed environment, a natural environment, a totally uniform background (e.g., dissection plate, paper) or a background containing a human body (even partially visible). |

**e. Automatic classification methods tested**

As a simple baseline method, the cultivation status was predicted based on the observer-provided iNaturalist "*cultivated*" tag (iNat_tag). This baseline enables us to evaluate the discriminatory power of this tag, which is decomposed in the confusion matrix of Appendix 4. Besides, we tested eleven classification algorithms to account for three potential problems in our dataset: the high number of input variables compared to the annotated/training sample size, the class imbalance with an over-representation of wild compared to cultivated records, and the classification uncertainty due to doubtful records. The eleven algorithms, based on two model families, logistic regression and random forests (Breiman, 2001), are described below per family and were implemented in R.

Each classification algorithm was fitted on eight combinations of input sets (resulting in 88 models) in order to compare the predictive power of the input sets and to compare models of increasing complexity: (i) basic input set only (basic), (ii) basic and iNaturalist specific input sets combined (inat), (iii) basic and geographic input sets (geo), (iv) basic and iNaturalist specific and geographic input sets (inat_geo), (v) basic and visual input sets (visu), (vi) basic, visual and geographic input sets (visu_geo), (vii) basic, iNaturalist specific and visual input sets (inat_visu), and (viii) all input sets combined (all).

The simplest algorithm (glm_full) was a binary logistic regression fitted using all variables of the input set. The model was trained only on manually annotated records classified as cultivated or wild, with doubtful records excluded. It was implemented with the glm function in R. A weighted version was implemented, where the log-likelihood weight of each sample depends on its annotated class so that weights of all records within a class summed to one (glm_full_weight). We implemented another version of the logistic regression integrating doubtful records as "soft" labels (glm_full_soft), namely assuming that the true class of a doubtful record is random with an equal probability of 0.5 for being cultivated or wild. We also combined these two approaches in a weighted soft-label model (glm_full_soft_weighted). We implemented an L1-penalised version of the logistic regression using the glmnet package (Friedman et al., 2010, Friedman et al., 2021), a method often used for input variable selection when facing many candidate variables with a limited sample size. The L1 penalty adds an extra cost to the negative log-likelihood which is proportional to a hyper-parameter $\lambda$>0 and to the sum of the absolute values of the model coefficients. As $\lambda$ increases, more coefficients of input variables are shrunk exactly to zero. We applied the L1-penalised logistic regression to the non-doubtful records only (algorithm: glmnet), optimising the value of $\lambda$ through cross-validation (cv.glmnet function), which served as a prior input variable selection step before applying a simple logistic regression to the selected input variables (glm_selec). Besides, we integrated the doubtful records to the two latter models as done previously with glm_soft, and these implementations are referred to as glmnet_soft and glm_selec_soft, respectively.

We considered three random forest classifiers implemented with the randomForest R package (Liaw and Wiener, 2002): A binary classification random forest trained only on cultivated and wild records (RF), another integrating doubtful records as a third class (RF_doubt), and a final version similar to the previous one, except that samples were weighted so that the weights of samples within a class summed to one hence accounting for class imbalance (RF_doubt_weighted). In all cases, the hyperparameters (mtry: number of input variables randomly evaluated at each branch, and nodesize: minimum size of terminal nodes) were first optimised to minimise the Out-Of-Bag error (OOB) before growing a random forest of 500 trees with the best hyper-parameters. In all cases, 500 trees were sufficient to stabilise the Out-Of-Bag error estimates.

For all algorithms and input sets, whenever a continuous predictor (dToCanalM, dToRiverM, tagFreq, positional_accuracy) was included in a model, its quadratic term was also included to allow for non-linear relationships with cultivation status.

### f. Evaluation

The predictive methods were primarily evaluated for their ability to predict non-doubtful records. We computed the predictive accuracy (overall frequency of correctly predicted status), sensitivity to the cultivated status (detection rate of cultivated records) and to the wild status (detection rate of wild records).

In addition to evaluating binary classification performance, we also evaluated the ability of the methods to account for uncertainty by predicting a third class, *doubtful*, when observations could not be reliably classified as either cultivated or wild (Figure 4). For the logistic regression model, which predicted only two classes (cultivated versus wild), a doubtful prediction was defined as a case where the predicted probability was close to 0.5. Then, we determined a symmetric margin around 0.5 such that all records whose predicted probabilities lay within the margin were predicted as doubtful. For instance, with a margin of 0.1, all records predicted as cultivated with a probability between 0.4 and 0.6 were classified as doubtful. We tested three margin values of 0.1, 0.2 and 0.4 and kept only 0.1 for the main analyses (Figure 4) as it maximised the mean predictive accuracy for all logistic models (see Figure S4.1 of the Appendix 3).

Out-of-sample predictive performance was evaluated with two types of cross-validation: (i) standard 20-fold cross-validation based on equal random folds whose results are presented in the main text and (ii) leave-one-species-out cross-validation, i.e. a 12-fold cross-validation where each fold corresponded to all records of one species, to evaluate the transferability of the method to species not seen in training as presented in Appendix 5. For the iNat_tag method, evaluation metrics were computed once on all records.

### g. Input variables importance

The importance of all input variables for prediction was assessed using the most complete and best-performing model, namely the RF_doubt fitted with all input variables. We used the variable importance framework of the randomForest R package which measures (i) the mean decrease in classification accuracy when the input variable values are randomly permuted in the OOB predictions, and (ii) the mean decrease of the Gini index averaged across branches where the input variable was chosen for the split and averaged across trees. The Gini index is the criterion used by the random forest algorithm to select predictor variables at each branch when constructing decision trees.

## 3. Results

**Table 2.** Number of iNaturalist records annotated manually for their cultivation status per species and the percentage of each status (wild, cultivated and doubtful). We systematically annotated all trusted records of the 12 species in the area, i.e. with sufficient taxonomic revision by the community and geolocation certainty, representing about 1.5% of all trusted records. The numbers provided between brackets include only Research Grade records (no inatTag).

| Species | Annotated records | % wild | % cultivated | % doubtful |
|---:|---:|:---:|:---:|:---:|
| *Melaleuca viminalis* | 76 | 43.4 | 14.5 | 42.1 |
| *Centranthus ruber* | 378 | 78.0 | 8.73 | 13.2 |
| *Duranta erecta* | 64 | 1.6 | 48.4 | 50.0 |
| *Echium candicans* | 45 | 57.8 | 20.0 | 22.2 |
| *Ficus carica* | 82 | 57.3 | 19.5 | 23.2 |
| *Hypericum canariense* | 87 | 89.7 | 1.2 | 9.1 |
| *Phoenix canariensis* | 137 | 12.4 | 72.3 | 15.3 |
| *Psidium guajava* | 64 | 9.4 | 78.1 | 15.5 |
| *Schinus terebinthifolia* | 313 | 45.7 | 18.8 | 35.5 |
| *Solanum lycopersicum* | 59 | 18.6 | 76.3 | 5.1 |
| *Spartium junceum* | 153 | 54.2 | 11.1 | 34.6 |
| *Tecoma stans* | 33 | 12.1 | 15.2 | 72.7 |
| **Total** | **1491 (1353)** | **50.0 (54.4)** | **25.2 (21.1)** | **24.8 (24.5)** |
| **Mean across species** | **124.3 (112.8)** | **40.0 (42.4)** | **32.0 (29.8)** | **28.0 (27.8)** |

iNaturalist records contained a substantial proportion of clearly cultivated plants (25.2%) in our manually annotated sample of 1,491 records, even when considering only Research Grade records (21.1%), i.e. without inatTag (**Table 2**). Furthermore, this proportion of cultivated records may under-represent the proportion for broader taxonomic groups as some species with a large part of the annotated records had much lower rates of cultivated records than other species (e.g. *Centranthus ruber*, 378 records, 8.73% cultivated), and the percentage of cultivated records climbed to 32% when taking the average across species (and 29.8% for Research Grade records only, **Table 2**). We also found that 24.8% of records had doubtful status, so that the real percentage of cultivated plants among annotated iNaturalist records could be as high as approximately 50%. We found drastic variations of the proportions of cultivated, wild and doubtful records across species, with up to 89.7% of wild records for *Hypericum canariense*, 78.1% of cultivated records for *Psidium guajava* and 72.7% of doubtful records for *Tecoma stans* **(Table 2)**.

**The iNat_tag was partly informative in discriminating between wild and cultivated records** (accuracy = 0.74, Figure 4-top). However, despite its very good sensitivity to wild records (99% of actual wild records are detected by the absence of tag; Figure 4-bottom), the sensitivity to the cultivated records was low (24%, Figure 4-middle). Actually, most records tagged as cultivated in iNaturalist were truly cultivated (91.8% without doubtful records, 65.2% with it, **Table S4.1**) but 21.4% of untagged records (Research Grade) were also cultivated, which is comparable to the total percentage of cultivated records (25.2%), meaning that the absence of tag gives little information about the cultivation status.

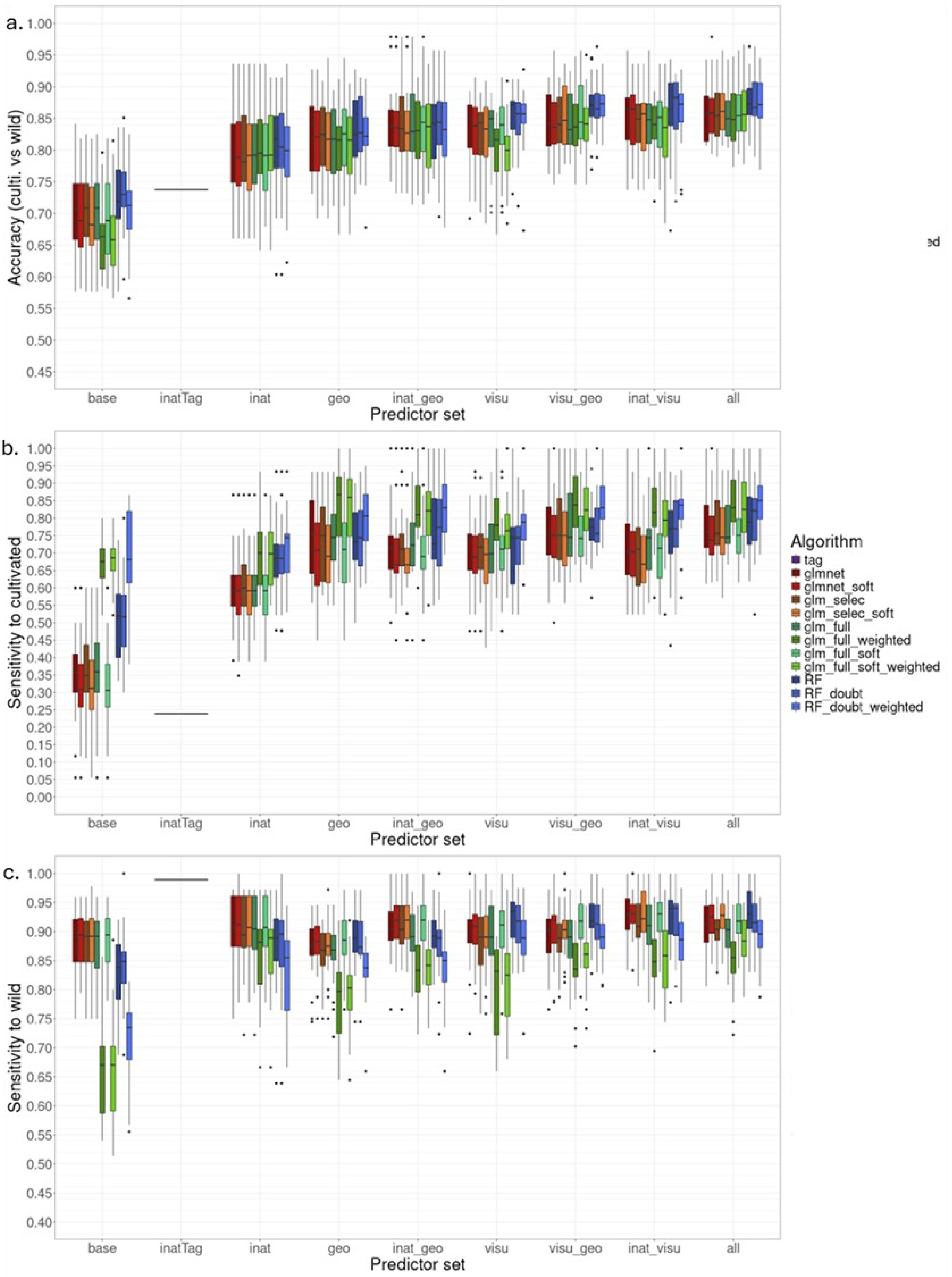

**Figure 4.** Boxplot of cross-validation (CV) performance metrics when classifying only wild versus cultivated records (n=1120) across the 20 CV folds for different input sets and classification algorithms. We display **a**, classification accuracy, **b**, sensitivity/recall for cultivated records, and **c**, sensitivity/recall for wild records.

**Geographic, visual or iNaturalist-specific predictors drastically improved the classification of the cultivation status compared to the inatTag or basic predictors alone**. Indeed, all algorithms performed better when using iNaturalist-specific, geographic or visual input sets compared to the basic input set only, and compared to the inatTag baseline (Figure 4). Indeed, for each algorithm, the three input sets (inat, geo and visu) improved their median classification accuracy in discriminating between wild and cultivated (Figure 4a) and the three classes (Figure 5) compared to the basic input set and the iNatTag method. The best algorithm based on basic input variables (RF) yielded a median two-class accuracy of 74%, similarly to the iNatTag method.

**Visual predictors alone predicted better cultivation than iNaturalist or geographic predictors alone, but were species specific.** Indeed, when using only one input set, visual (visu) predictors were better than iNaturalist (inat) predictors under random data splits, as the best visu algorithm had a median accuracy of 0.871, i.e. 0.054 higher (+6.6%) than the best inat algorithm, and the average accuracy across all visu algorithms was also higher by 0.03 (+3.7%) as illustrated by Figure 4-top. Nevertheless, visual inputs alone were less transferable across species than iNaturalist inputs, as shown by the leave-one-species-out CV experiment (Figure S5.1 of Appendix 5) in which visu algorithms had an overall average accuracy 0.014 (1.8%) lower than inat algorithms, and the best visu algorithm was 0.007 (1%) lower than the best inat algorithm. The comparison was qualitatively the same between visual and geographic (geo) algorithms, but geo algorithms were overall better than inat algorithms and much more transferable across species than visu algorithms with an increase of 0.019 (+2.3%) in best algorithm accuracy.

**Combining two input sets (inat, geo and visu) brought complementary information and more consistent and transferable predictions across species.** Indeed, predictor sets combining several predictor types (e.g. inat_visu combines visual and iNaturalist-specific predictors) tended to improve the median accuracy compared to the predictor sets with only one predictor type (inat and visu predictor sets), which was observed when classifying records as cultivated versus wild (**Figure 4a.**) and when accounting for doubtful records (**Figure 5**). For instance, RF achieved an 84% median accuracy in classifying cultivated versus wild records with inat_geo predictors (**Figure 4a.**) while it achieved a median accuracy of 82.5% with geo predictors and 81% with inat predictors. However, there was little or no improvement when combining all three predictor types compared to pair-sets (inat_visu, inat_geo, visu_geo) (**Figure 4a**). A similar pattern was observed for the leave-one-species-out cross-validation scheme (**Figure S5.1 of Appendix 5**).

**Random forest algorithms using all input variables yielded overall the best performance.** The best two-class accuracy was obtained with the RF algorithm using all input variables, with a mean accuracy of 87.9% (median of 88.2%, Figure 4-top), which was very closely followed by RF_doubt also using all predictors, with a mean accuracy of 87.6% (median of 88%, Figure 4-top). The best three-class accuracy was obtained with RF_doubt on the visu_geo variable set with a mean accuracy of 70.6% (median of 70.7%). Besides, RF_doubt_weighted using all predictors yielded the best mean two-class accuracy for the transferability across species (leave-one-species-out cross-validation scheme, **Figure S5.1 of Appendix 5**) with 84.7%.

**The prevalence of wild records resulted in a low sensitivity to cultivated records, except for weighted algorithms.** Due to the larger prevalence of wild records compared to cultivated records, the sensitivity to wild records (their rate of detection) was overall higher than that for cultivated records for most algorithms (**Figure 4b. and c.**). However, the weighted algorithms (glm_full_weighted, glm_full_soft_weighted and RF_doubt_weighted) partly corrected for this bias, as they had a higher sensitivity to cultivated records than their non-weighted counterparts, and a lower sensitivity to wild records (**Figure 4b. and c.**).

**Prediction accuracy remained high for held-out species.** Model performance declined slightly overall but remained quite high when evaluated for their taxonomic transferability based on a leave-one-species-out cross-validation experiment (**Appendix 5, Figure S5.1**). Indeed, the overall mean two-class accuracy across all models and input sets for the leave-one-species-out cross-validation experiment was 77.3% versus 81.8% for the 20-fold cross-validation experiment (**Figure 4**). The prediction performance across species was also visibly more variable than for random folds when comparing inter-quartile ranges of Figure S5.1 and Figure 4.

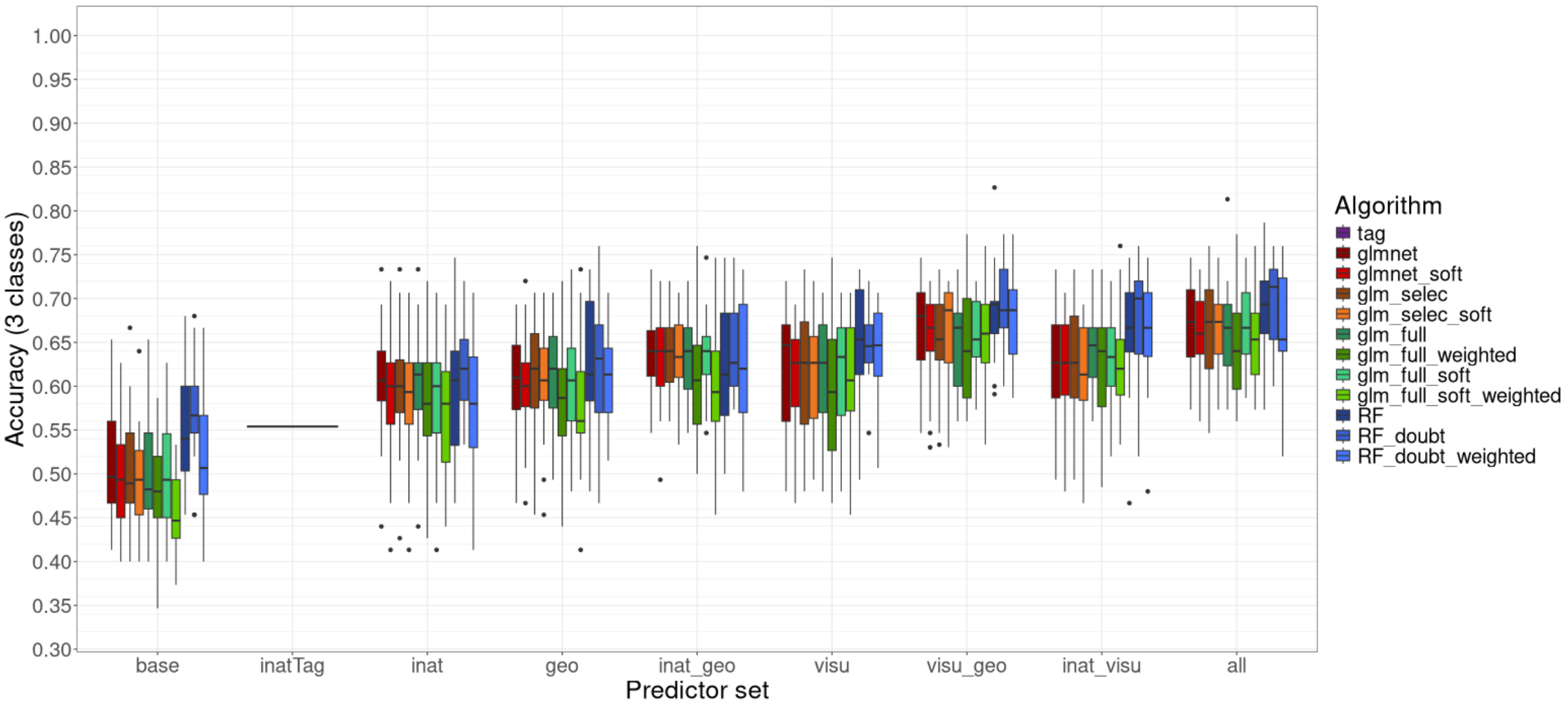

**Figure 5.** Boxplot of cross-validation (CV) accuracy when classifying all three classes (n=1491) across CV-blocks for different sets of predictors and classification algorithms.

**Geographic and visual predictors were the most important, followed by some iNaturalist-specific ones.** We plot ranked predictor importances in the RF_doubt algorithm using all predictors as input (one of the best models overall) for both importance metrics considered, i.e. the decrease in accuracy and in Gini index when permuting the given predictor (**Figure 6**). The most important predictor according to both metrics was the anthropized aspect of the visual context (visu_ctx_anthropized), and the second was the indication that the record was in a protected area (protecArea). Besides, geographic and visual predictors appeared overall as the most important type of predictors. Indeed, 4 geographic predictors were in the top 10 most important for both metrics, while for visual predictors, 3 predictors were in the top 10 for the decrease in accuracy and 4 were in the top 10 for the decrease in Gini index. The important visual predictors were all tied to the context (picture background). iNaturalist predictors appeared slightly less important as none of them were in the top 3 for both metrics, but 3 were in the top 10 for the decrease in accuracy, and 2 in the top 10 for decrease in Gini index. These important iNaturalist specific predictors included the iNatTag and the frequency of records tagged as cultivated by the observer (tagFreq).

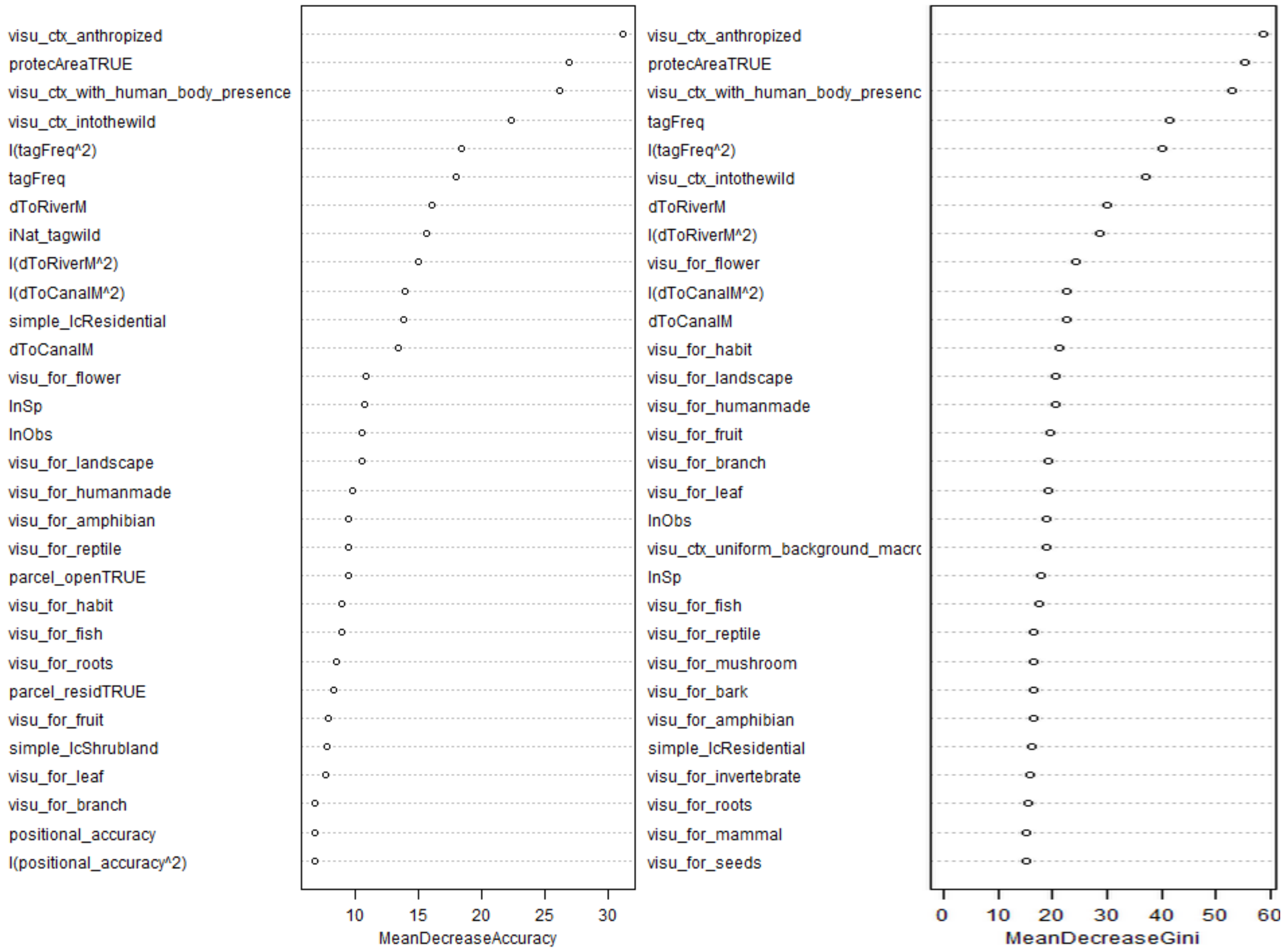

**Figure 6.** Predictors ordered by decreasing importance in the random forest fitted to all annotated records (RF_doubt) using all predictors. Importance is estimated with two metrics: the mean decrease in classification accuracy (left panel), and the mean decrease of the Gini index (right panel).

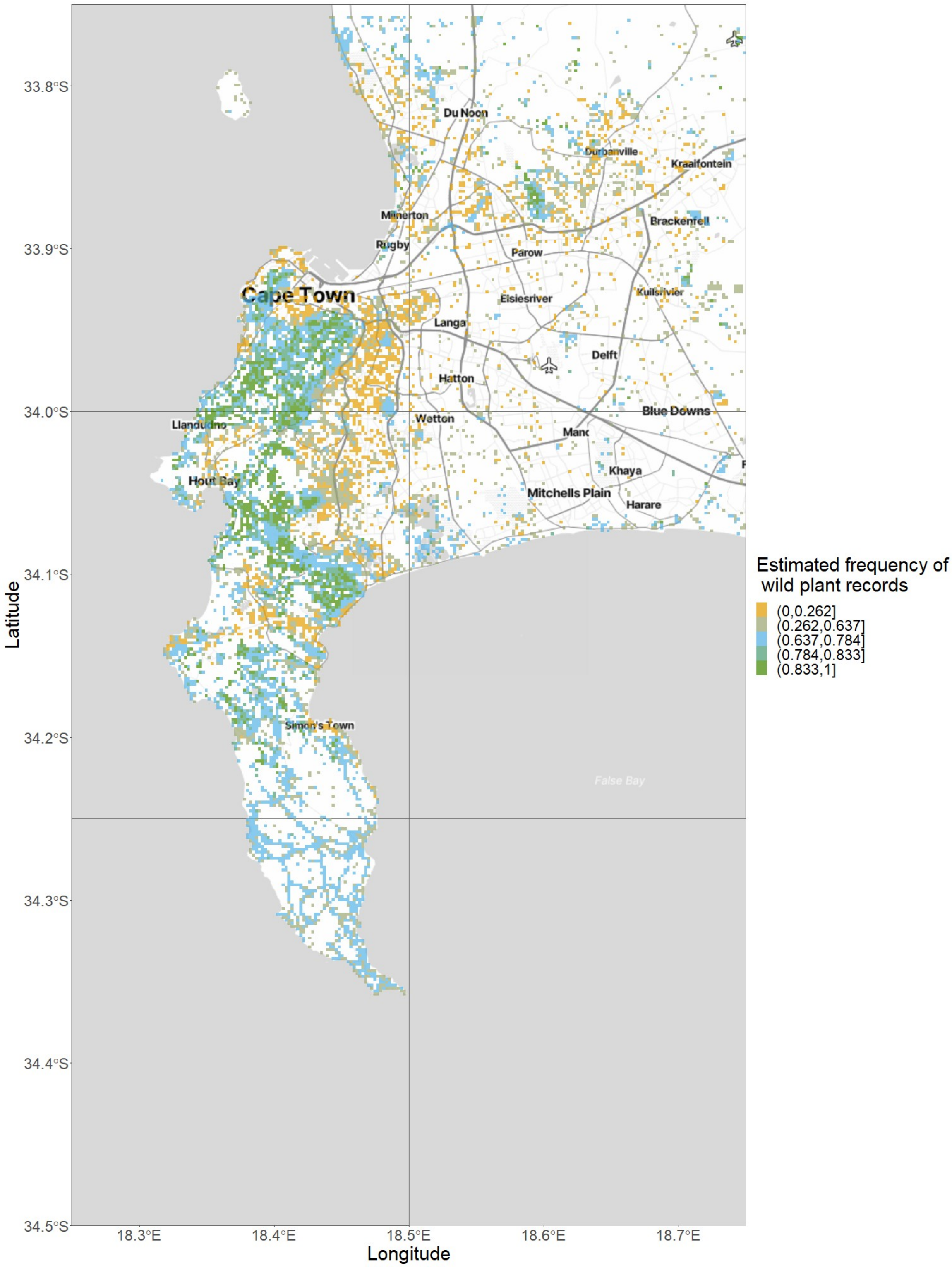

**Figure 7.** Map of estimated frequency of wild plant records at 200 m resolution based on our best model predictions (RF_doubt_weighted) applied to the 99,408 trusted iNaturalist records of all plant species. This frequency is the local proportion of actual iNaturalist records classified as wild by the model in each cell.

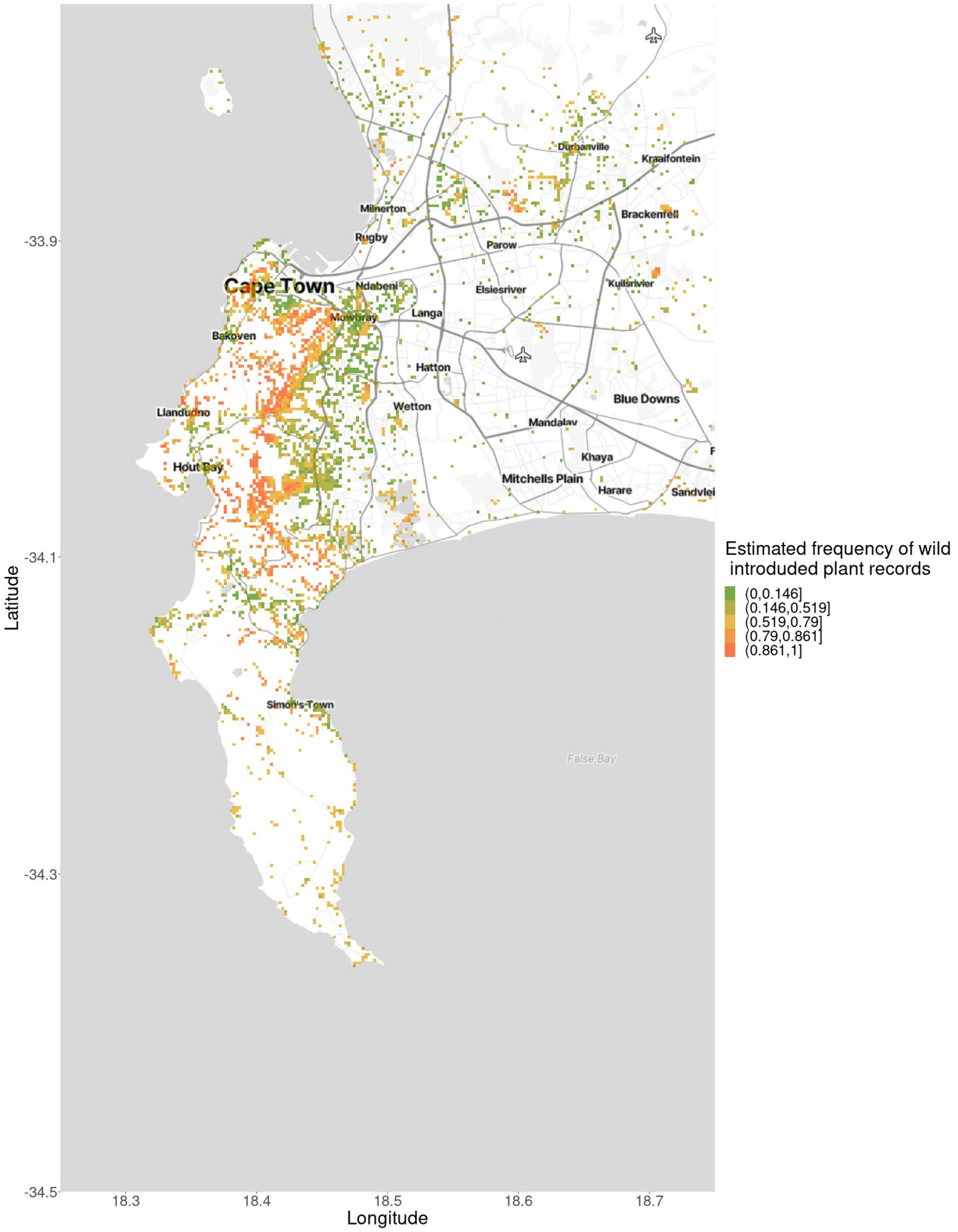

**Figure 8.** Map of estimated frequency of wild introduced plant records at 200 m resolution based on our best model predictions (RF_doubt_weighted) and computed specifically on the records of the 529 species introduced to the area (Rebelo et al., 2025). This frequency is the local proportion of records of introduced plants that are classified as wild by the model, which can be seen as a prediction of the potential naturalisation level of the plants observed in iNaturalist in each cell.

To illustrate potential final uses of the method, we used the algorithm RF_doubt_weighted trained on the 1,491 iNaturalist records manually annotated for the cultivated status of only 12 species to predict the cultivated status of the 99,408 trusted records of all plant species around Cape Town. We mapped these predictions in (i) frequency of records predicted as 'wild' per 200m cell considering all observations (**Figure 7**), and (ii) the frequency of records predicted as 'wild' but considering only records of introduced species **(Figure 8)**, thus excluding native plants, which could inform on the local frequency of potentially naturalized alien plants. These Figures are overall coherent with the geography of the area. They show an overall large portion of records predicted as wild in the south of the Cape peninsula, hosting the Table Mountain National Park, and even more wild records on Table Mountain, south of Cape Town and north of the peninsula. In contrast, **Figure 7** shows a very low proportion of records predicted as wild in the suburbs of Cape Town, which are indeed poor in natural areas. Besides, **Figure 8** points out areas with particularly high proportions of wild records among introduced plants. This Figure highlights that introduced plants are mostly cultivated along the main road from Cape Town city to its southern suburban expansion while the belt separating this road from the slopes of the Table Mountain National Park is the main interface of spread towards the natural environment and a known naturalisation hotspot. Then, potential naturalisation hotspots are notably located on the northern and eastern flanks of Table Mountain, at the interface between natural and urban areas.

# 4. Discussion

**A decision framework to classify citizen science plant records as cultivated or wild.** We use a definition of the cultivation status at the plant specimen level that can help to anticipate naturalisations. This definition does not require the plant to have been planted, sown or propagated deliberately by humans, but focuses on the active selection or promotion of an existing plant by humans to the detriment of other plants that would have otherwise competed with it for natural resources. Then, a wild alien plant record means that the species is at least a casual alien (sensu Richardson et al., 2000): it survives outside cultivation, but it is not necessarily yet able to maintain a self-sustaining population without repeated introduction (i.e. the naturalisation stage). In other words, the species is at least at stage C1 or higher in the framework of Blackburn et al. (2011). We further proposed an evidence-based decision framework to determine the cultivation status of a plant specimen from the information contained in its digital iNaturalist record, when this information is sufficient. Following this framework, we manually annotated the cultivation status of 1,491 trusted iNaturalist plant records covering 12 alien invasive species around Cape Town, hence creating the first dataset of this kind to our knowledge.

**Contamination of iNaturalist by cultivated specimens and the "cultivated" tag.** The analysis of the annotated records showed that contamination of iNaturalist records by cultivated plants is common and must be taken into account when considering using these data. However, contamination varied across the 12 alien invasive species that were closely inspected in our study likely reflecting a variable invasive potential among plants used for ornamental and horticultural purposes. Systematic approaches are needed to address data contamination by cultivated plants as simply restricting iNaturalist data to Research Grade records does not sufficiently remove cultivated records, with the potential of reducing the quality of the data for ecological analyses that focus on wild observations (Richardson and Potgieter, 2024). There is potential for the user-provided tag (iNatTag) in identifying records of cultivated plants. However, its effectiveness is limited where cultivated records are not tagged. Indeed, in the majority of cases where the

iNatTag was absent, it did not inform about the cultivation status, and its use was also inconsistent across species (cf Appendix 5) resulting in an overall poor detection power for cultivated plants.

**Automating cultivation status assessment from plant records.** In order to provide a rapid alternative to human annotation or a complementary preliminary screening tool, we assessed the potential for automating the prediction of the cultivation status using machine learning methods. Machine learning methods achieved a good overall performance, as our best model classified cultivated and wild specimens with 87% accuracy, which was much better than relying on the user-provided tag for cultivation from iNaturalist (74%). The best classification models relied strongly on information about the geographic context and visual features from the image posted on iNaturalist. It is well known that under class imbalance, as is the case here, the fitted models show a bias towards the majority class, lowering the ability to detect the minority class (He and Garcia, 2009, Benkendorf et al., 2023). Indeed, in our case, employing weighted classification algorithms, with both logistic regression and random forest models, addressed data imbalance towards wild records and reduced overpredicting the wild status. Performance in our models also improved to an extent when richer predictors (iNaturalist-specific, geographic and visual) were added to the basic and iNatTag predictors in the models. Performance tended to improve when combining diverse predictors with the best overall accuracy obtained when all predictors were added. Integrating more accurate and informative predictors (e.g. satellite imagery) could potentially further make such predictions more reliable. Yet, this would likely require a vast amount of annotated observations covering more diverse species and contexts in order to avoid overfitting, while the cost of manual annotation remains relatively high.

**Transferability.** Our transferability experiment showed that our best algorithms yielded almost as good predictions when applied to new species, unseen during model training, with about 85% accuracy in distinguishing cultivated and wild specimens. Yet, this evaluation may be positively biased by the chosen species pool, where all species are well known invasive species occurring as cultivated and wild in the area. Addressing the transferability to species rarely observed as cultivated or wild may be more challenging. Some of our best algorithms could be directly transferred to observations from other large citizen science platforms (e.g. Pl@ntNet, observation.org), because the iNaturalist specific predictors (such as the “cultivated” tag) did not bring a clear performance gain to the models in addition to visual and geographic predictors. In this study, we did not assess the extent to which the approach could be transferred to other spatial contexts, and our results are restricted to the habitats found around Cape Town. Nevertheless, this could be addressed in the future through manual annotation for a reasonable global sample of records in a well-chosen stratified set of locations.

**Limits of our approach.** Our criteria for the manual annotation of the cultivated and wild status were tailored for urban and peri-urban contexts. They were generally conservative, potentially leading to an overclassification of the “doubtful” status, but simplified to avoid adding specific conditions to firmly decide on the status if these conditions are only relevant in rare cases. However, this may lead us to overlook important groups of contexts which might justify a revision of these criteria in the future. Future revisions could incorporate a broader range of land-use contexts which may improve the detection of early stage naturalisation. Such refinements would enhance the generalisability and scalability of the framework, ultimately improving its applicability across regions. For instance, our criteria should be used with caution in other land use contexts such as forests. In exploited forests, where the production cycle lasts for decades, it is common to see spontaneous bushes or trees growing freely close to the exploited species, which does not mean that this competition has not been actively eliminated in past years.

Another limitation is that we used cross-validation for model comparison, but did not perform an independent test of final performance, as is generally recommended (see e.g. Tredennick et al., 2021). Indeed, it seemed problematic here to set aside a significant sample as a test set, given the already limited number of annotated records, particularly for certain species (e.g. *Tecoma stans*). Consequently, our best models' performance estimates may be slightly inflated due to the model selection step. Moreover, the validation scheme employed here does not account for spatial autocorrelation, potentially leading to overly optimistic performance estimates (Dormann et al., 2007; Tredennick et al., 2021). Future studies should therefore incorporate spatially structured validation or independent test datasets to provide a more robust assessment of model transferability and predictive reliability.

The limited capacity to annotate large numbers of records for cultivation status is currently preventing the assessment of the approach to a larger diversity of spatial contexts and taxa. This limitation is particularly important given the reliance of supervised machine-learning models on high-quality labelled data, which directly influences model performance and transferability. Addressing this challenge could require a stratified sampling strategy to select a parsimonious number of new records to annotate while maximising their coverage over new contexts. Besides, developing more efficient annotation strategies, such as active learning, semi-supervised approaches or transfer learning (Bothmann et al. 2023) would maximise the utility of the mass of available data while reducing manual effort.

**Limits of opportunistic observations.** The large number of untagged cultivated plants in our iNaturalist dataset reflects a common issue also reported in other studies (López-Guillén et al. 2024) where iNaturalist users fail to indicate whether an observation is wild or cultivated, even when it is clearly cultivated. While our manual annotation showed a more precise picture of the cultivation status, a substantial part of the records had a doubtful status. The cultivation status is typically ambiguous in some situations. For instance, close-up photos, trees identified by flowers or fruits whose pictures generally only show sky in the background (e.g. *Psidium guajava* in our data) or plants growing within cultivated stands (e.g. along edges or fences). It is intrinsically more difficult to tell the cultivation status of certain species. For instance, slow maturing species for which flowers are key for identification are particularly difficult to classify, as it is often no longer possible to tell whether the plant was cultivated in earlier stages. To minimise intrinsic ambiguity of the observations, we suggest citizen science platforms (e.g. Pl@ntNet, iNaturalist, observation.org, naturgucker) to encourage observers to take one picture showing the whole plant within its context, including the ground surrounding its main stem or trunk, a perspective that can be critical to detect cultivated specimens (López-Guillén et al. 2024; Richardson and Potgieter, 2024).

**End uses of cultivation status predictions.** The absence of a tag or precise spatial context, such as a picture showing the surrounding habitat, can present challenges in assessing the naturalisation stage of alien plants in large cities such as Cape Town with a high concentration of non-native species (Kowarik, 2011; Gaertner et al. 2016) and where alien plants frequently occur at the interface between gardens and adjacent natural areas (Alston and Richardson, 2006; Afonso et al. 2020; Potgieter et al. 2024). Categorising plants as cultivated or wild is also a crucial step in many ecological analyses (Richardson and Potgieter 2024). Ecological niche models and habitat suitability models (Guisan et al., 2017), aiming at capturing species' potential abiotic preferences, are notoriously biased by the contamination of records by many cultivated plant specimens, as is commonly the case with crowdsourced records (Botella et al., 2018). In this context, niche models will overestimate the range of suitable conditions of a species because many specimens are present out of their niche through nurturing (e.g., fertilisation, irrigation, indoor planting). This may also affect policy decisions, including the listing of invasive taxa in risk assessments or the assessment of threatened species (Blackburn et al., 2014; Rodrigues et al., 2006). For instance, overestimating a species' future range under global change scenarios (Beninde et al., 2023; López-Guillén et al., 2024) may generate false alarms regarding potential invasions, or alternatively lead to unwarranted reassurance about species that are actually in decline. Thus, our method could typically improve the precision of models that predict the potential area of plant invasion (Ibanez et al., 2009) by removing cultivated plants from the records on which these models are fitted. Although our findings highlight the critical need to distinguish wild from cultivated observations to avoid bias in ecological analyses, cultivated records themselves remain valuable and should not be dismissed outright. Indeed, high levels of propagule pressure (e.g., the repeated introductions of the same alien species at a location) also increase the likelihood of successful establishment (Potgieter and Cadotte 2020). For example, the horticultural industry has been a major pathway for the repeated importation of many alien plant species into urban areas (Padayachee et al. 2017, Van Kleunen et al., 2018) and the escape of ornamental alien plants from cultivation in gardens has resulted in some of the most extensive biological invasions (Dehnen-Schmutz et al., 2007; Geerts et al., 2013, 2017; Holmes et al., 2018). As a result, alien plant species make up a large proportion of urban floras relative to native species (Pyšek, 1998; Afonso et al., 2020; Potgieter and Cadotte, 2020; Potgieter et al., 2024).

**Conclusion.** Capitalising on the mass of plant records from the citizen science platform iNaturalist, our study proposed evidence-based decision criteria to manually classify the observed plants as cultivated, wild or doubtful. We also showed a proof of concept that machine learning can automate the classification process at scale, with (i) good performance, (ii) good transferability across taxa despite a small performance loss, (iii) potential transferability to similar data platforms like Pl@ntNet and to other regions subject to further validation. Yet, assessing the broader relevance of this approach was difficult given the limited capacity to manually annotate the cultivation status of records for a wide range of regions, ecological contexts and taxa. Hence, the proposed framework has considerable potential to support the assessment of the naturalisation status of alien plant species, identify invasion hotspots, or sanitise ecological analyses currently contaminated by the inclusion of many cultivated records.

## Acknowledgements

The R script and data to reproduce Tables and Figures are available on our Github repository: https://github.com/ChrisBotella/cultivated_vs_wild_inat/. The background geographic maps of Figures 1, 7 and 8 are based on Stadia Maps, Stamen Design, OpenMapTiles and contributions to OpenStreetMap. We thank Dr. Mlu Nsikani for his advice and insights. We declare the use of generative AI solely for text editing. The authors reviewed and edited the content as needed and take full responsibility for the content of the published article. C.B. and C.H. were supported by the National Research Foundation of South Africa (grant 89967). C.H. was also supported by the European Union's Horizon Europe Research and Innovation Programme (No. 101059592; B3 –

Biodiversity Building Blocks for Policy). S.G. was supported by the National Research Foundation of South Africa (Grant CSRP23041994970). DMR and LJP were supported by the Centre for Invasion Biology and the Department of Botany and Zoology, Stellenbosch University. LJP also received support from the Natural Sciences and Engineering Research Council of Canada (grant 386151). The views and opinions expressed are those of the authors and do not necessarily reflect those of the funding bodies. Neither the European Union nor the funding agencies can be held responsible for them.

# Supplementary material

## Appendix 1: Land parcel data

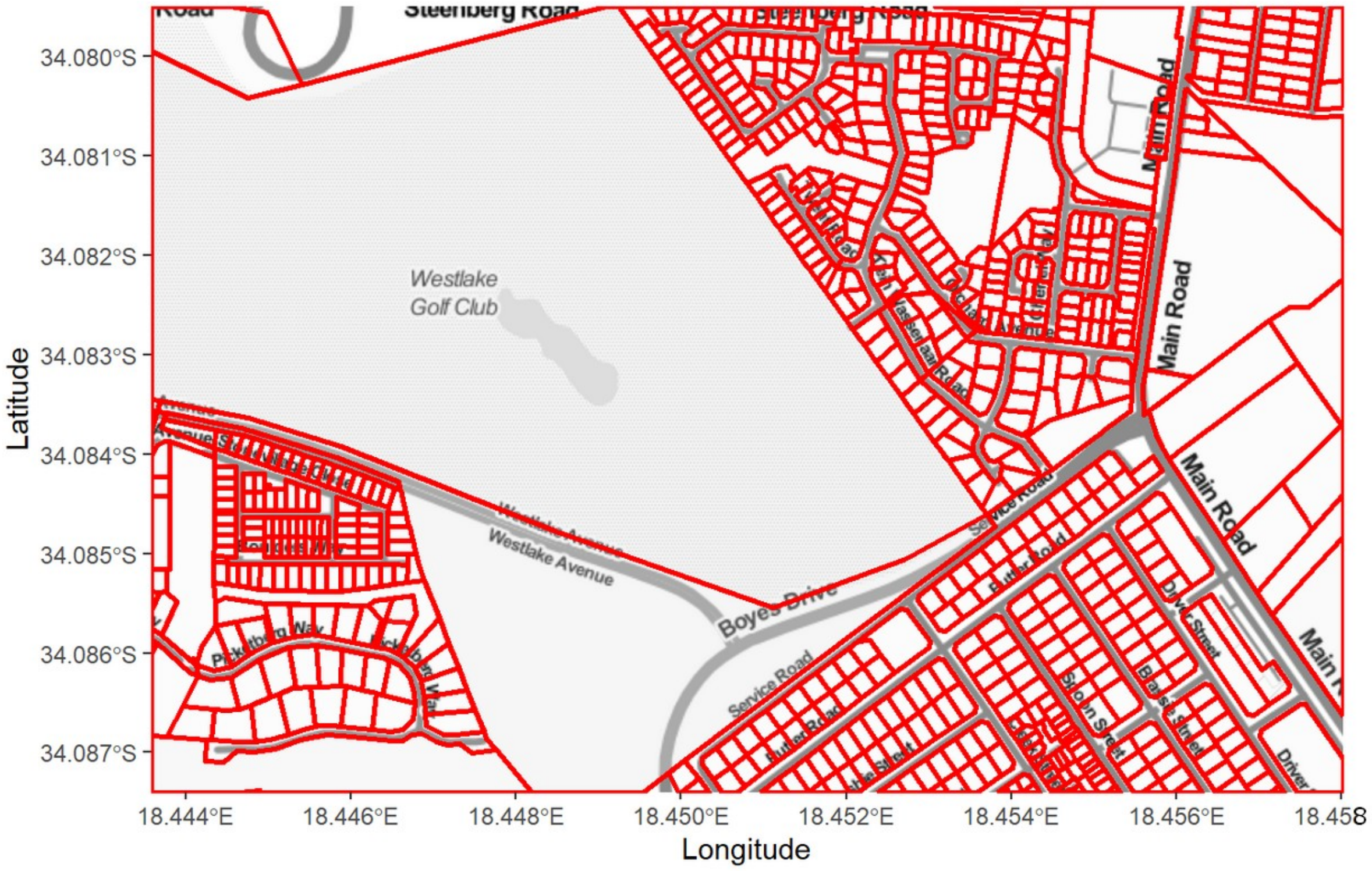

**Figure S1.1.** Illustration of the land parcel polygons (solid red lines) in the Westlake suburb of Cape Town showing polygons of roads, residences and a park.

## Appendix 2: Simplified land cover classes versus Globcover classes

| Raster value | GlobCover description | simple_lc |
|---|---|---|
| 0 | | Water |
| 1 | contiguous (indigenous) forest | Natural Wooded Land |
| 2 | contiguous low forest & thicket | Natural Wooded Land |
| 3 | dense forest & woodland | Natural Wooded Land |
| 4 | open woodland | Natural Wooded Land |
| 5 | contiguous & dense plantation forest | Planted Forest |
| 6 | open & sparse plantation forest | Planted Forest |
| 7 | temporary unplanted (clear-felled) plantation forest | Planted Forest |
| 8 | low shrubland (other) | Shrubland |
| 9 | low shrubland (fynbos) | Shrubland |
| 10 | low shrubland (succulent karoo) | Shrubland |
| 11 | low shrubland (nama karoo) | Shrubland |
| 12 | sparsely wooded grassland | Other |
| 13 | natural grassland | Other |
| 14 | natural rivers | Other |
| 15 | natural estuaries & lagoons | Other |
| 16 | natural ocean & coastal | Water |
| 17 | natural lakes | Water |
| 18 | natural pans (flooded @ observation times) | Other |
| 19 | artificial dams (including canals) | Other |
| 20 | artificial sewage ponds | Other |
| 21 | artificial flooded mine pits | Other |
| 22 | herbaceous wetlands (currently mapped) | Other |
| 23 | herbaceous wetlands (previously mapped) | Other |
| 24 | mangrove wetlands | Other |
| 25 | natural rock surfaces | Other |
| 26 | dry pans | Other |
| 27 | eroded lands | Other |
| 28 | sand dunes (terrestrial) | Other |
| 29 | coastal sand & dunes | Other |
| 30 | bare riverbed material | Other |
| 31 | other bare | Other |
| 32 | cultivated commercial permanent orchards | Other |
| 33 | cultivated commercial permanent vines | Other |
| 34 | cultivated commercial sugarcane pivot irrigated | Other |
| 35 | cultivated commercial permanent pineapples | Other |
| 36 | cultivated commercial sugarcane non-pivot | Other |
| 37 | cultivated emerging farmer sugarcane non-pivot | Other |
| 38 | commercial annual crops pivot irrigated | Other |
| 39 | commercial annual crops non-pivot irrigated | Other |
| 40 | commercial annual crops rain-fed / dryland | Other |
| 41 | subsistence / small-scale annual crops | Other |
| 42 | fallow land & old fields (trees) | Other |
| 43 | fallow land & old fields (bush) | Other |
| 44 | fallow land & old fields (grass) | Other |
| 45 | fallow land & old fields (bare) | Other |
| 46 | fallow land & old fields (low shrub) | Other |
| 47 | residential formal (tree) | Residential |
| 48 | residential formal (bush) | Residential |
| 49 | residential formal (low veg / grass) | Residential |
| 50 | residential formal (bare) | Residential |
| 51 | residential informal (tree) | Residential |
| 52 | residential informal (bush) | Residential |
| 53 | residential informal (low veg / grass) | Residential |
| 54 | residential informal (bare) | Residential |
| 55 | village scattered (bare & low veg/ grass combo) | Residential |
| 56 | village dense (bare & low veg / grass combo) | Residential |
| 57 | smallholdings (tree) | Urban Vegetation |
| 58 | smallholdings (bush) | Urban Vegetation |
| 59 | smallholdings (low veg / grass) | Urban Vegetation |
| 60 | smallholdings (bare) | Built-up |
| 61 | urban recreational fields (tree) | Urban Vegetation |
| 62 | urban recreational fields (bush) | Urban Vegetation |
| 63 | urban recreational fields (grass) | Urban Vegetation |
| 64 | urban recreational fields (bare) | Urban Vegetation |
| 65 | commercial | Built-up |
| 66 | industrial | Built-up |
| 67 | roads & rails (major linear) | Built-up |
| 68 | mines: surface infrastructure | Other |
| 69 | mines: extraction pits, quarries | Other |
| 70 | mines: salt mines | Other |
| 71 | mine: tailings and resource dumps | Other |
| 72 | land-fills | Other |
| 73 | fallow land & old fields (wetlands) | Other |

**Table S2.1.** The 73 Globcover land cover classes and the seven simplified land cover classes used in our models to predict the cultivation status.

# Appendix 3: Optimal margin to predict doubtful records with logistic models

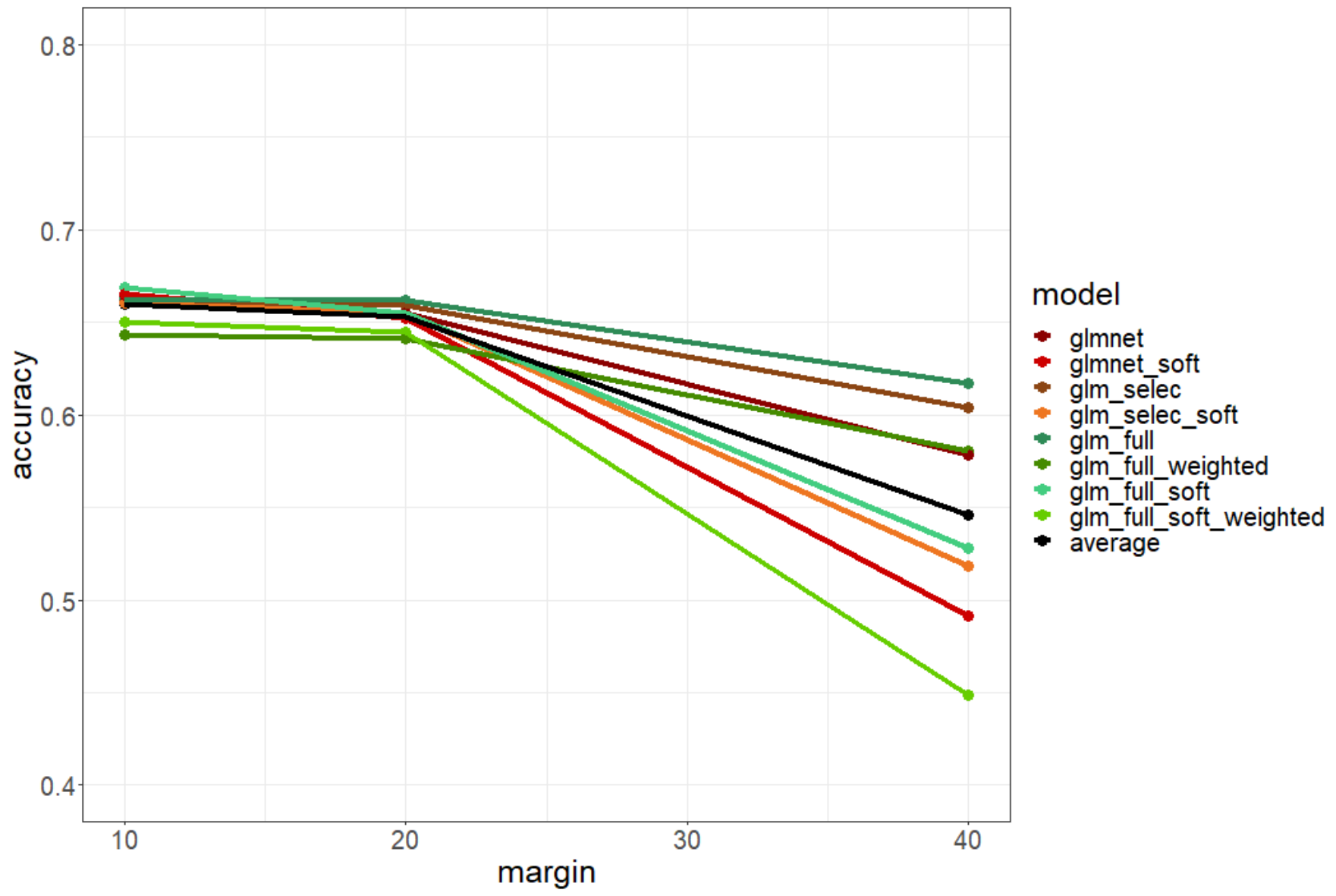

**Figure S3.1.** Mean predictive accuracy of logistic models for predicting the three classes (cultivated, wild and doubtful), where the doubtful class is determined by a specified probability margin (100x). The 0.1 probability margin maximised the predictive accuracy of all models.

## Appendix 4: Confusion matrix of iNatTag and cultivated status

| | Cultivated | Wild | Doubtful | **Total** |
|---|---|---|---|---|
| Reported “cultivated” | 90 | 8 | 40 | **138 (9.3%)** |
| No tag | 286 | 736 | 331 | **1353 (90.7%)** |
| **Total** | **376 (25.2%)** | **744 (49.9%)** | **371 (24.9%)** | |

**Table S4.1.** Number of annotated records per annotated cultivation status and reporting as “captive/cultivated” in iNaturalist.

## Appendix 5: Transferability of models across species

We evaluated the robustness and taxonomic transferability of the classification models to other species than the one annotated by performing a leave-one-species-out cross-validation on the annotated records. For each left-out species among the twelve annotated species, we fitted each model (combination of algorithm and predictor set) on the records of the eleven others, predicted on the records of left-out species and measured the same evaluation metrics as elaborated in the main text. The number of records per species varied from 33 to 378, hence leaving between 75% and 98% of the total annotated records for model fitting depending on the left-out species. The distribution of classification accuracy in classifying cultivated versus wild classes and in three classes (including “doubtful”) are plotted for all models in **Figure S5.1.**

As expected, both metrics were overall lower than in the 20-fold cross-validation experiment (blind split, see manuscript), with a best mean accuracy of 84.5% on two classes (vs 87.9% on the blind split) and a best mean accuracy of on three classes of 63.2% (vs 70.6%), showing that it is more difficult to generalise to new species than to a new record from a species that was present in model training. Even though the inatTag method performed better in median accuracy on two classes than it did in the blind split experiment, it also had a tremendous variability across left-out species, showing that the reporting of the cultivated status is inconsistent across alien species. The performance rank of predictor sets and models were rather similar to the blind split experiment.

Despite the overall loss of performance compared to the blind split experiment, the best models had informative predictions. In terms of mean two-class-accuracy, the best model was **RF_doubt_weighted** fitted on **all predictors** (84.5%). In terms of mean three-class-accuracy (when accounting for doubtful records) the best model was once again RF fitted on the combination of visual and iNaturalist specific predictors (inat_visu), with 63.2%, closely followed by **RF_doubt_weighted** fitted on **all predictors** (62.5%).

Weighted algorithms correcting for class-imbalance (specifically RF_doubt_weighted) were often more successful in taxonomic transferability than algorithms that did not account for it. Four out of the top-five models in terms of two-class-accuracy and two out of the top-five in terms of three-class accuracy were based on RF_doubt_weighted. This result is coherent with the different profiles of class imbalance encountered across species, with some species having mostly cultivated records while others had mostly wild records, and it showed the importance of correcting for this bias to generalise better.

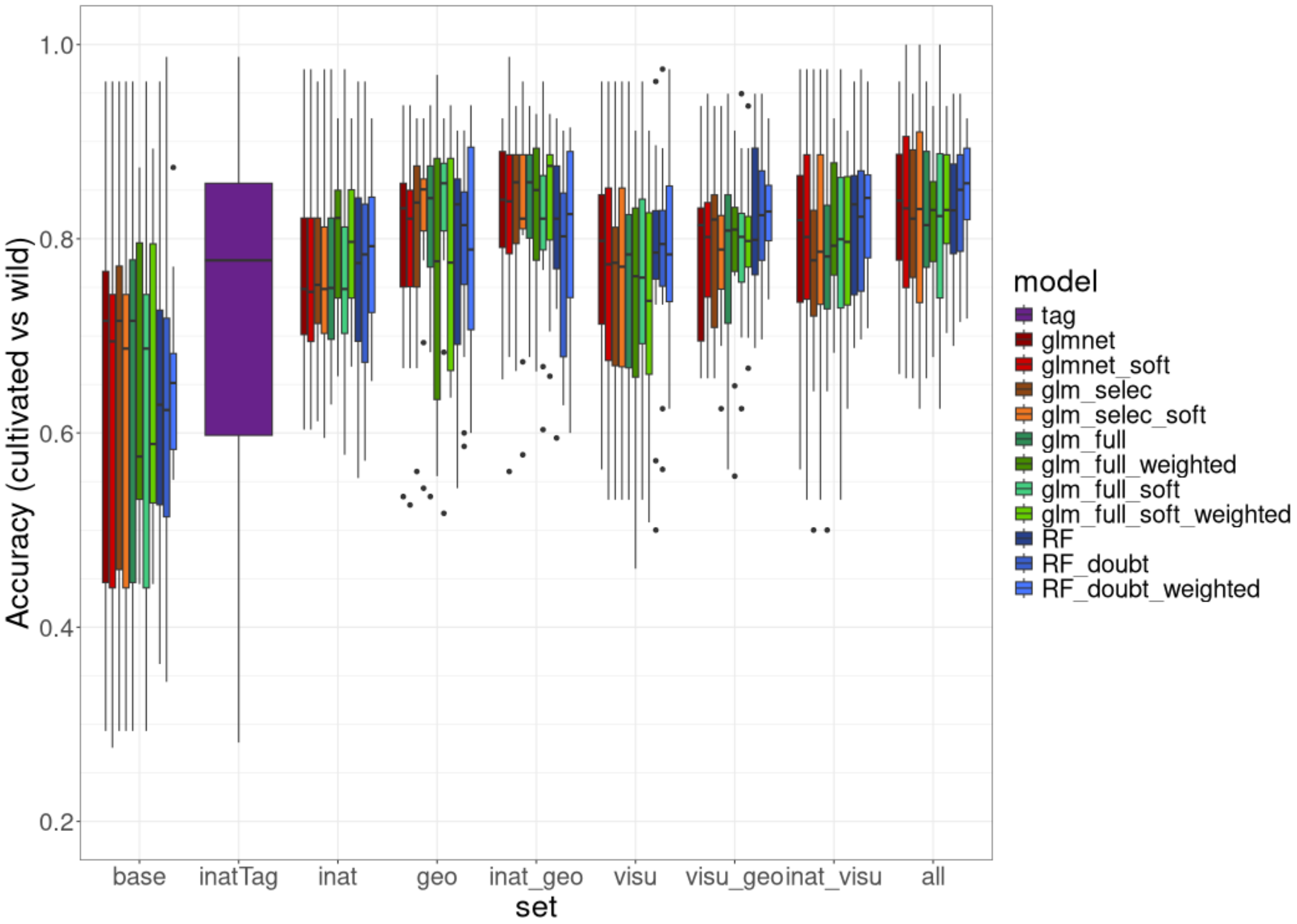

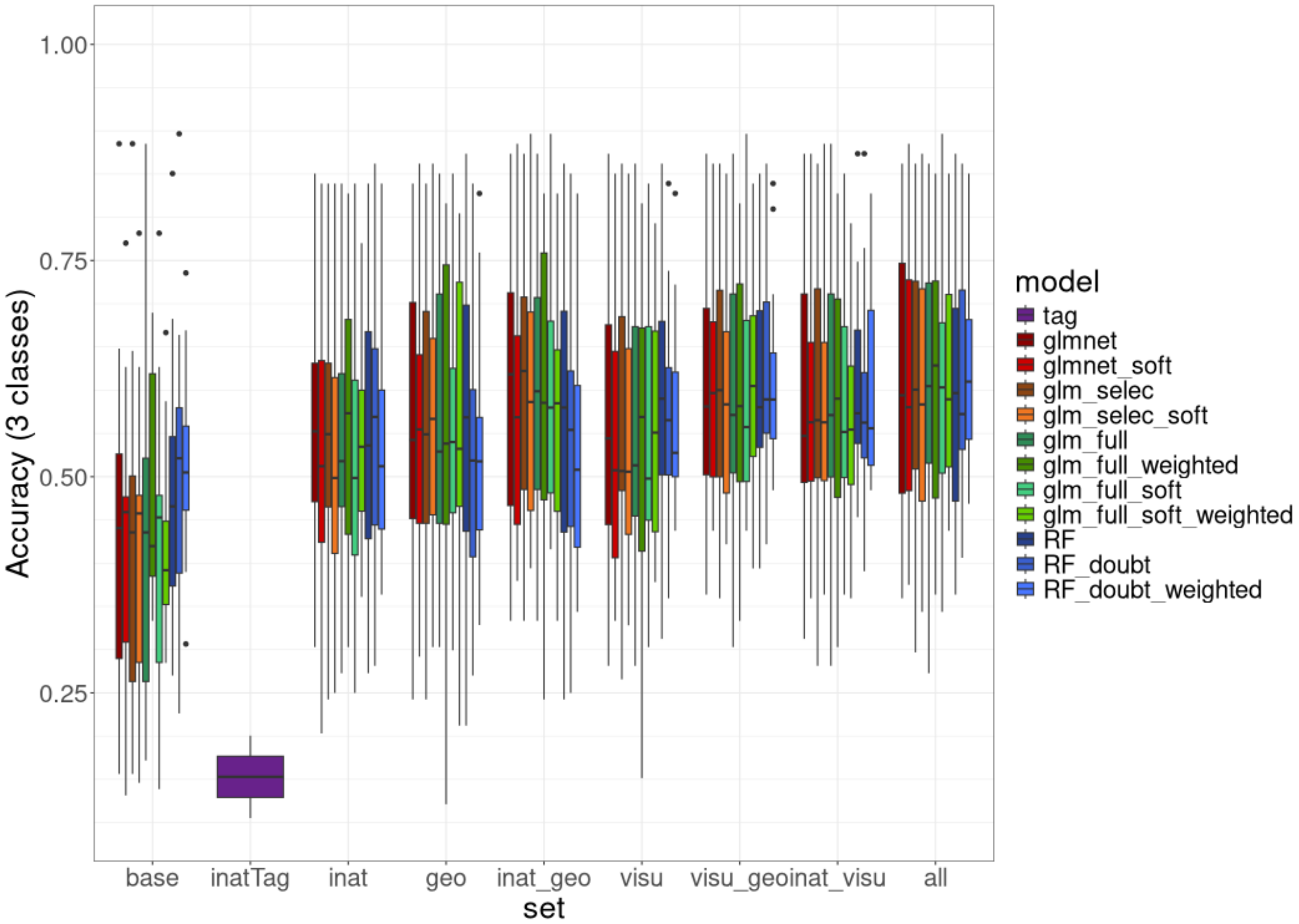

**Figure S5.1.** Taxonomic transferability of the 89 classification methods assessed with a leave-one-species-out cross-validation procedure on the annotated records of 12 alien species.